\documentclass[aip,aapm,mph,bm,twocolumn,english,aps,prl,superscriptaddress,groupedaddress]{revtex4}
\usepackage[utf8]{inputenc}
\usepackage{amsmath}
\usepackage{amssymb}
\usepackage{graphicx}
\usepackage{esint}
\usepackage{booktabs}
\usepackage{multirow}

\makeatletter
\@ifundefined{textcolor}{}
{%
 \definecolor{BLACK}{gray}{0}
 \definecolor{WHITE}{gray}{1}
 \definecolor{RED}{rgb}{1,0,0}
 \definecolor{GREEN}{rgb}{0,1,0}
 \definecolor{BLUE}{rgb}{0,0,1}
 \definecolor{CYAN}{cmyk}{1,0,0,0}
 \definecolor{MAGENTA}{cmyk}{0,1,0,0}
 \definecolor{YELLOW}{cmyk}{0,0,1,0}
}

\usepackage{epsfig}
\usepackage{epstopdf}
\usepackage{bm}
\usepackage{epsfig}\usepackage{mathrsfs}
\usepackage{dcolumn}
\usepackage{color}
\usepackage{miniplot}
\usepackage{minitoc}
\usepackage{blindtext}
\@ifundefined{showcaptionsetup}{}{%
\PassOptionsToPackage{caption=false}{subfig}}

\def\be{\begin{equation}}
\def\ee{\end{equation}}
\def\bea{\begin{eqnarray}}
\def\eea{\end{eqnarray}}

\allowdisplaybreaks[2]
\DeclareUnicodeCharacter{00A0}{~}

\usepackage{babel}

\usepackage{babel}

\makeatother

\usepackage{babel}
\begin{document}

\preprint{APS/123-QED}
\title{Unification of valley and anomalous Hall effects in a strained lattice}
\date{\today}
\author{Jiale Yuan}
\affiliation{Interdisciplinary Center for Quantum Information and State Key Laboratory of Modern Optical Instrumentation,Zhejiang Province Key Laboratory of Quantum Technology and Device and Department of Physics,
Zhejiang University, Hangzhou 310027, China}
\author{Han Cai}
\affiliation{Interdisciplinary Center for Quantum Information and State Key Laboratory of Modern Optical Instrumentation,Zhejiang Province Key Laboratory of Quantum Technology and Device and Department of Physics,
Zhejiang University, Hangzhou 310027, China}
\author{Congjun Wu}
\affiliation{School of Science, Westlake University, Hangzhou 310024, China}
\author{Shi-Yao Zhu}
\affiliation{Interdisciplinary Center for Quantum Information and State Key Laboratory of Modern Optical Instrumentation,Zhejiang Province Key Laboratory of Quantum Technology and Device and Department of Physics,
Zhejiang University, Hangzhou 310027, China}
\author{Ren-Bao Liu}
\affiliation{Department of Physics, Centre for Quantum Coherence, and The Hong Kong Institute of Quantum Information Science and Technology, The Chinese University of Hong Kong, Shatin, N. T., Hong Kong, China}
\author{Da-Wei Wang}
\email{dwwang@zju.edu.cn}
\affiliation{Interdisciplinary Center for Quantum Information and State Key Laboratory of Modern Optical Instrumentation,Zhejiang Province Key Laboratory of Quantum Technology and Device and Department of Physics,
Zhejiang University, Hangzhou 310027, China}
\affiliation{CAS Center for Excellence in Topological Quantum Computation, University of Chinese Academy of Sciences, Beijing 100190, China}

\begin{abstract}
Two dimensional lattices are an important stage for studying many aspects of quantum physics, in particular the topological phases. The valley Hall and anomalous Hall effects are two representative topological phenomena. Here we show that they can be unified in a strained honeycomb lattice, where the hopping strengths between neighboring sites are designed by mimicking those between the Fock states in a three-mode Jaynes-Cummings model. Such a strain induces an effective magnetic field which results in quantized Landau levels. The eigenstates in the zeroth Landau level can be represented by the eigenstates of a large pseudo-spin. We find that the valley Hall current and the chiral edge current in the Haldane model correspond to the spin precession around different axes. Our study sheds light on connection between seemingly unrelated topological phases in condensed matter physics.
 \end{abstract}

\maketitle

Lattice dynamics beyond real space has been investigated in cold atoms \cite{Cooper2019} and photonic systems  \cite{Yuan2018, Ozawa2019} with synthetic extra dimensions. By using the Fock states of multiple cavities coupled to a two-level atom, lattices in arbitrary dimensions can be synthesized \cite{Wang20151,Wang20152,Wang2016, nwaa196}. These Fock-state lattices offer a new tool for engineering quantum states (such as the preparation of mesoscopic superposition states \cite{Wang2016}) and investigating high dimensional topological physics \cite{nwaa196}. In particular, the inhomogeneous coupling strengths between the Fock states, which are proportional to $\sqrt{n}$ with $n$ being the photon number in the 
Fock states, resemble strain in real-space lattices. Such a specific strain creates unconventional edges and provides a unique platform for investigating exotic phenomena in strained lattices \cite{Su1979,Suzuura2002,Morozov2006,Morpurgo2006,Manes2007,Pereira2009,Pikulin2016,Guan2017,Jia2019}. 

In graphene, a nonuniform strain can induce an effective magnetic field with a strength much larger than that of a real magnetic field \cite{Castro2009,Goerbig2011,guinea2010,Levy2010,Yan2013}. Similar mechanisms have been implemented in atomic and molecular \cite{gomes2012,tian2015}, photonic \cite{Rechtsman2013} and phononic \cite{Yang2017,Wen2019} lattices to achieve quantized Landau levels. The strain induced magnetic field has opposite signs in the two valleys of graphene, which offers new controlling knobs for valleytronic engineering \cite{Zhai2010,Chaves2020,Schaibley2016,Pearce2016,Zhai2020,Etienne2020}. However, only excitations near the two valleys in weak strain field can be investigated with perturbative methods \cite{Suzuura2002,Manes2007,guinea2010}. The relation between the two valleys under the same strain, in particular for those that can open a band gap \cite{Pereira2009}, is still unclear. 

In this Letter, by mimicking the exactly solvable Fock-state lattices, we unify the valley Hall current \cite{xiao2007,yao2008,xiao2012} and the chiral edge current of the Haldane model \cite{haldane1988} in a strained tight-binding lattice. The strain results in quantized Landau levels. We use the eigenstates in the zeroth Landau level to construct a large pseudo-spin. The valley Hall current and the chiral edge current correspond to the spin precession about different axes. We also establish a mapping between the Aharonov-Bohm (AB) phase of an electron \cite{Aharonov1959} and the Berry phase of a spin \cite{berry1984}. This study sheds new light on hidden relations between different topological phenomena in condensed matter physics.

\begin{figure}[t]
\centering
\includegraphics[scale=0.22]{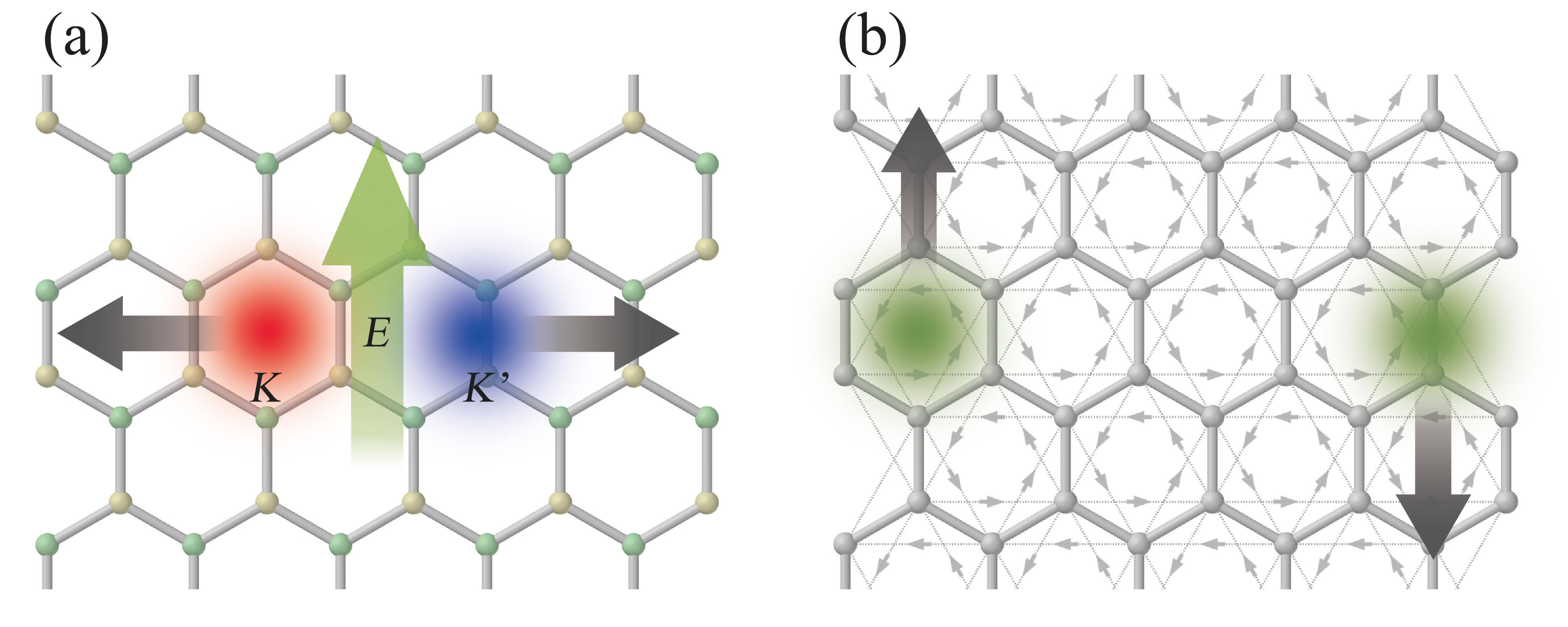}
\caption{(a) Valley Hall current.~In a honeycomb lattice with inversion symmetry breaking, the two valleys ($K$ and $K^\prime$) have opposite Berry curvatures. Wavepackets at the two valleys drift in opposite directions (as shown by the gray arrows) perpendicular to the subjected electric field (the big green arrow). (b) Haldane edge current. With phased next-nearest-neighbor hopping terms (the arrows on the dashed lines show the hopping directions attached with phase factor $i$) that break the time-reversal symmetry, edge currents propagate in opposite directions (the gray arrows).}
\label{fig0}
\end{figure}

We schematically show the essence of the valley Hall system and the Haldane model in Fig.~\ref{fig0}. In a honeycomb lattice where the inversion symmetry is broken, e.g., by different on-site potential of the two sublattices or a nonuniform strain field, the Berry curvatures at the two valleys have opposite signs. An electric field would induce Hall currents proportional to the Berry curvatures, such that the wavepackets at the two valleys drift in opposite directions \cite{xiao2007,yao2008,xiao2012}. On the other hand, in the Haldane model the next-nearest-neighbor (NNN) hopping terms break the time-reversal symmetry and the wavepackets propagate in opposite directions at the two edges \cite{haldane1988,Jotzu2014,Ho2017}. There is a dual relation between the electric field in the valley Hall system and the NNN terms in the Haldane model. In the valley Hall effect, the electric field breaks the inversion symmetry, and the valley currents at the two time-reversal symmetric points ($K$ and $K^\prime$) are opposite. In the Haldane model, the NNN terms break the time-reversal symmetry, and the resulted chiral edge currents at the inversion symmetric points (upper and lower edges) are opposite. Since the time-reversal symmetry is equivalent to momentum-reversal symmetry, such a dual relation indicates that these two effects are unified in an integrated phase space of position and momentum. We found that both of them can be represented by the precession of a large pseudo-spin. The states at the $K$ and $K^\prime$ points correspond to the pseudo-spin states at the north and south poles of the Bloch sphere. The Haldane edge states correspond to the pseudo-spin coherent states at the equator. The valley Hall current and the Haldane edge current correspond to the spin precession around a pseudo-magnetic field in and perpendicular to the lattice plane, respectively. 

\begin{figure}[t]
\centering
\includegraphics[scale=0.4]{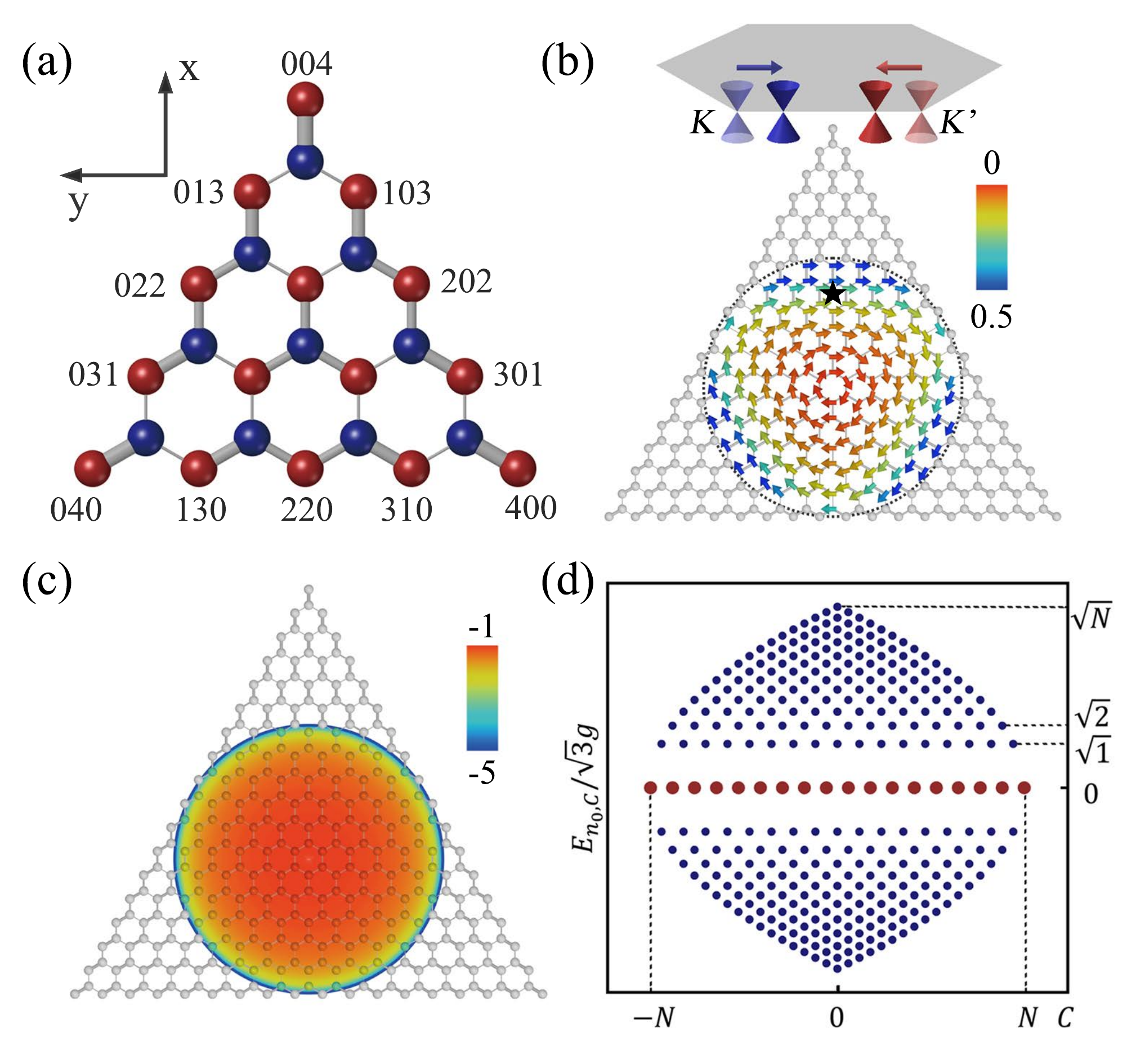}
\caption{Strain-induced effective magnetic field and Landau levels in a tight-binding honeycomb lattice. (a) A honeycomb tight-binding lattice with Hamiltonian in Eq.~(\ref{h}). Red and blue sites denote $a$ and $b$ sublattices. Numbers label the indices $pqr$ of $a$-sublattice sites on the boundary. The line widths are proportional to the hopping coefficients. (b) Displacements of the Dirac cones in the Brillouin zone (at the position denoted by ``$\star$'') and the distribution of the strain-induced vector potential $A^+(\mathbf{r})/q_0$, where $q_0=4\pi/3\sqrt 3 u$. The orientations of the arrows denote the directions and the colors denote the values. (c) Distribution of the strain-induced effective magnetic field $B^+(\mathbf{r})/B_0$. (d) The eigenenergy spectrum of the tight-binding lattice, $E_{n_0,C}^\pm=\pm\sqrt{3n_0} g$ for the upper and lower $n_0$th Landau level and chirality $C$.}
\label{fig1}
\end{figure}

To demonstrate that relation, we construct a tight-binding model of exactly flat bands for arbitrary size of lattices \cite{Rachel2016,Ilan2020,Li2013,Li2012} by borrowing the coupling strengths in the Fock-state lattice of a multi-mode Jaynes-Cummings (JC) model \cite{jcmodel,Wang2016,nwaa196},
\begin{equation}
H_{JC}=g\sum_{i=1}^3 (c_i \sigma^++\sigma^-c_i^\dagger),
\label{hjc}
\end{equation}
where $g$ is the vacuum Rabi splitting, $c$ and $c^\dagger$ are the bosonic annihilation and creation operators of the photons in three cavities, and $\sigma^+\equiv |b\rangle\langle a|$ and $\sigma^-\equiv |a\rangle\langle b|$ are the raising and lowering operators of a two-level atom with the excited state $|b\rangle$ and ground state $|a\rangle$. The Fock states with $p$, $q$ and $r$ photons and the atom in the state $|a/b\rangle$ is denoted as $|b/a;p,q,r\rangle$. The Hamiltonian $H_{JC}$ conserves the total excitation number $N=p+q+r+(\sigma_z+1)/2$. 
By replacing the Fock states $|b;p-1,q,r\rangle$ and $|a;p,q,r\rangle$ with lattice sites, we construct a tight-binding lattice Hamiltonian, 
\begin{equation}
H=\sum_{pqr}g(\sqrt{p}b^\dagger_{p-1jk}+\sqrt{q}b^\dagger_{pq-1k}+\sqrt{r}b^\dagger_{pqr-1})a_{pqr}+h.c.,
\label{h}
\end{equation}
where $a_{pqr}$ and $b_{pqr}$ are the fermionic annihilation operators on the two sublattices. The honeycomb lattices contains $(N+1)^2$ sites with a triangular boundary, as shown in Fig.~\ref{fig1} (a). We meet the boundary when one of $p,q,r$ becomes zero. We define the center of the triangular boundary as the origin and the $x$-axis pointing at the top vertex and the $y$-axis pointing horizontally to the left, as shown in Fig.~\ref{fig1} (a). The coordinates of each site $pqr$ are $x=u(2r-p-q)/2$ and $y=\sqrt{3}u(p-q)/2$, which can be obtained by seeking the expectation values of the operators $x\rightarrow u(2c_{3}^{\dagger}c_{3}-c_{1}^{\dagger}c_{1}-c_{2}^{\dagger}c_{2})/{2}$ and $y\rightarrow {\sqrt{3}u}(c_{1}^{\dagger}c_{1}-c_{2}^{\dagger}c_{2})/2$ in the corresponding Fock-state lattice. Here ``$\rightarrow$'' means a mapping to the equivalent quantities in the Fock-state lattice.

In the adiabatic limit, the lowest order Hamiltonian can be expanded near the Dirac points \cite{Pereira2009,Goerbig2011,Juan2012},
\begin{equation}
H=\xi v_F[\mathbf{p}-e\mathbf{A}^\xi(\mathbf{r})]\cdot\sigma,
\end{equation}
where $v_F=gu\sqrt{3N}/2\hbar$ is the Fermi velocity with $u$ being the lattice constant, $\mathbf{\sigma}=\sigma_x\hat{x}+\sigma_y\hat{y}$ is the two-dimensional Pauli matrix vector, $\mathbf{p}$ is the lattice momentum,  $e$ is the electric charge, $\mathbf{A^\xi(\mathbf{r})}$ is the strain-induced gauge fields with $\xi=+$ and $-$ for the $K$ and $K^\prime$ points, and $\mathbf{r}$ is the displacement from the geometric center of the lattice. We obtain $\mathbf{A}^\xi(\mathbf{r})$ by calculating the displacements of the two Dirac points due to the strain \cite{Suzuura2002,Manes2007,Pereira2009} (see Fig.~\ref{fig1} (b)). The strain-induced effective magnetic field is (see Supplementary material),
\begin{equation}
\mathbf{B}^\xi(\mathbf{r})=\nabla \times \mathbf{A}^\xi(\mathbf{r})=-\xi\frac{B_0}{\sqrt{1-\mathbf{r}^2/R^2}}\hat{z},
\label{magnetic}
\end{equation}
where $B_0=2\hbar/Neu^2$, and $R\equiv Nu/2$ is the radius of the incircle of the triangular lattice boundary. By substituting the lattice constant of a real graphene $u=0.14$ nm, we obtain $B_0=6.5\times 10^4/N$ tesla. For a piece of graphene with micrometer size, $N$ is in the order of $10^4$ and $B_0$ is a few tesla, which is achievable in distorted graphene \cite{guinea2010}.~Obviously, when $N\rightarrow \infty$, the effective magnetic field at the center of the lattice $B_0\rightarrow 0$. The effective magnetic field is undefined on the incircle $|\mathbf{r}|=R$ (see Fig.~\ref{fig1} (c)), where the two Dirac points merge and a band gap opens \cite{Pereira2009}. A Lifshitz topological phase transition \cite{Lifshitz1960} occurs on the incircle, which is an unconventional edge that encloses a piece of semimetal in real space and connects the $K$ and $K^\prime$ points in momentum space. Such an edge enables us to integrate the valley Hall system and the Haldane model in a unified picture.

We diagonalize $H$ in Eq.~(\ref{h}) by borrowing coefficients of the eigenstates of $H_{JC}$ in Eq.~(\ref{hjc}), which can be easily diagonalized by introducing three combinational modes  \cite{nwaa196,Wang2016},
$d_0^\dagger=\sum_j c_j^\dagger/\sqrt{3}$ and $d_\pm^\dagger=\sum_j c_j^\dagger \exp{(\pm2ij\pi/3)}/\sqrt{3}$ (see Supplementary material), and the eigenenergy spectrum is shown in Fig.~\ref{fig1} (d). It is obvious that the eigenenergies scale with $\sqrt{n}$, the same as the Landau levels of electrons in a magnetic field near the Dirac cones of graphene. There are $N+1$ zero-energy states,
$|\psi_{0,C}\rangle= |a;0,n_+,n_-\rangle_d$,
where $C=n_+-n_-$ and $n_{j}=\langle d_j^\dagger d_j\rangle$ in the state ket $\left|...\right\rangle_d$ is the photon number in the Fock state of the $d_j$ mode ($j=0,\pm$). By introducing the Schwinger representation of an angular momentum, $J_{x}=(d_{+}^{\dagger}d_{-}+d_{-}^{\dagger}d_{+})/2$, $J_{y}=i(-d_{+}^{\dagger}d_{-}+d_{-}^{\dagger}d_{+})/2$ and $J_z=(d_+^\dagger d_+-d_-^\dagger d_-)/2$ \cite{Sakurai2017}, we construct the eigenstates of a pseudo-spin $j=N/2$ by using the states $|\psi_{0,2m}\rangle$ with $m=-N/2, -N/2+1, ..., N/2$ being the magnetic quantum numbers. It is interesting to note that $J_x$ and $J_y$ are time-reversal even while $J_z$ is time-reversal odd, which turns out to be responsible for the symmetry properties in the valley and anomalous Hall effects, respectively.

\begin{figure}[ht]
\centering
\includegraphics[scale=0.4]{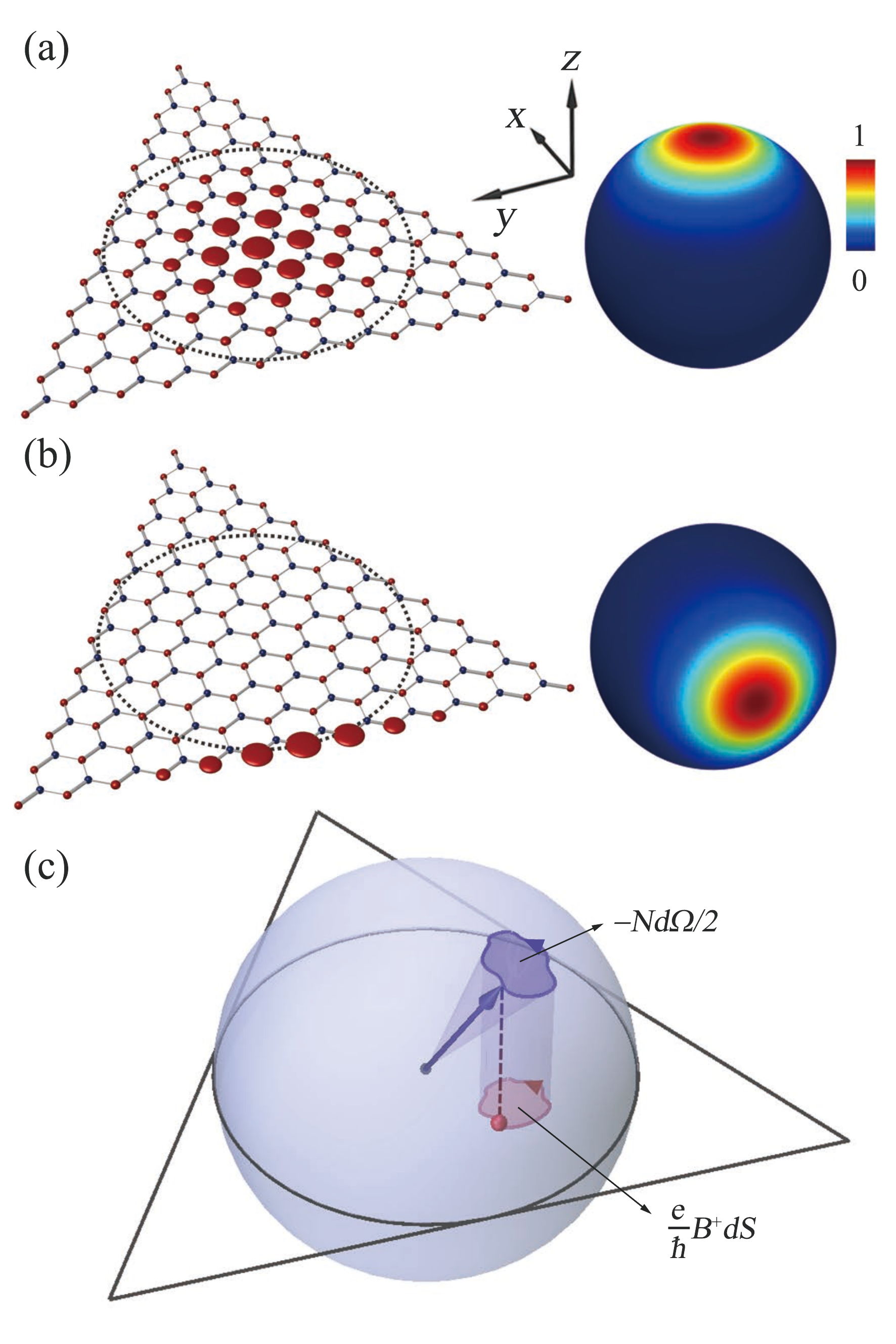}
\caption{States and Berry phases in the zeroth Landau level compared with those of a pseudo-spin. Quasidistribution function $W(\theta, \phi)=|\langle \theta^\prime, \phi^\prime|\theta, \phi\rangle|^2$ of the spin coherent states $|\theta,\phi\rangle=|0,0\rangle$ (a) and $|\pi/2,\pi\rangle$ (b) are shown with the probability distributions of the corresponding states $\psi_{\theta,\phi}(\mathbf{r})$ in the zeroth Landau level of the strained lattice. The radii of the filled circles in the lattice are proportional to the probabilities on the corresponding sites. (c) Relation between the Berry phase of the pseudo-spin and the AB phase in the the lattice, both accumulated in closed loops.}
\label{fig2}
\end{figure} 

\begin{figure*}[ht]
\centering
\includegraphics[scale=0.19]{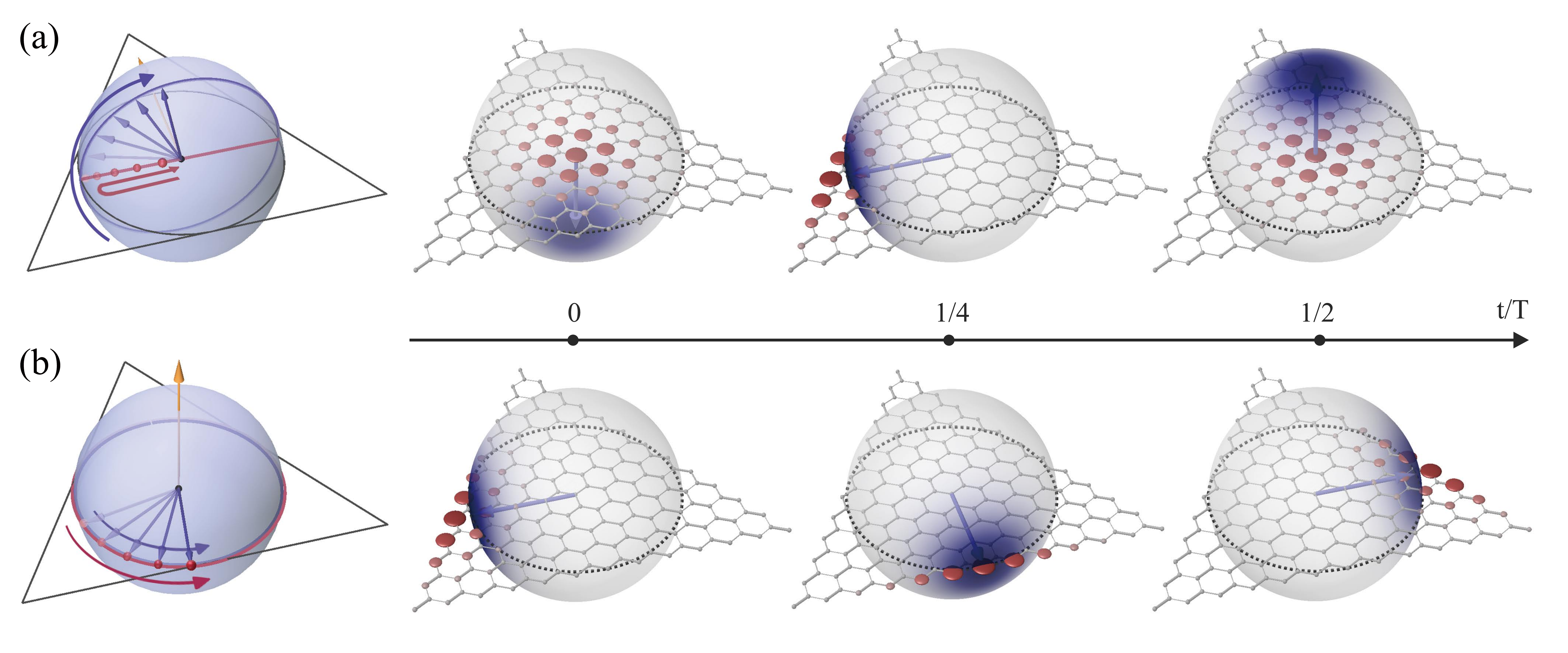}
\caption{A unified picture of electronic responses in the valley and anomalous Hall effects. (a) Time evolution of the electronic wave packet in a static electric field in $-{x}$ direction with the initial state $|\theta,\phi\rangle=|\pi,0\rangle$. (b) Time evolution of the edge state in the Haldane model with the initial state $|\theta,\phi\rangle=|\pi/2,\pi/2\rangle$. Figures on the left show the traces of the states during the time evolution, with the yellow arrows being the precession axes. The arrows to the Bloch sphere represent the spin coherent states. The curled blue and red arrows show the traces of the spin coherent state and the zeroth Landau level state, respectively. Figures on the right are three snapshots of the states during the evolution in sequence, at $t=0,T/4$ and $T/2$ with $T$ being the precession period. }
\label{fig3}
\end{figure*}

A pseudo-spin coherent state $|\theta,\phi\rangle$ \cite{Layton1990,Agarwal1997} (see Supplementary material)
corresponds to a wavefunction in the honeycomb lattice, $\psi_{\theta,\phi}(\mathbf{r})$, as shown in Fig.~\ref{fig2} (a) and (b). We establish the correspondence between the polar coordinates $(\theta, \phi)$ and the position $\mathbf{r}$ by overlapping the equator of the Bloch sphere and the incircle of the tight-binding lattice, as shown in Fig.~\ref{fig2} (c). The expectation value of the position $\mathbf{r}$ of the electronic wavefuntion $\psi_{\theta,\phi}(\mathbf{r})$ is the projection of the Bloch sphere point $(\theta,\phi)$ on the equator plane, i.e., $\langle \mathbf{r}\rangle=R\sin\theta\cos\phi\hat{x}+R\sin\theta\sin\phi\hat{y}$ (see Supplementary material). We adiabatically move the state $|\theta,\phi\rangle$ along a small loop that subtends a solid angle $d\Omega$ (see Fig.~\ref{fig2} (c)), the accumulated Berry phase is \cite{berry1984,Layton1990},
\begin{equation}
d\gamma=-Nd\Omega/2.
\label{gm1}
\end{equation}
The mapping between the pseudo-spin coherent state and the wavefunction ensures that $\psi_{\theta,\phi}(\mathbf{r})$ accumulates the same geometric phase during the adiabatic transport. In this case it is the AB phase,
\begin{equation}
d\gamma=\frac{e}{\hbar}B^\xi dS.
\label{gm2}
\end{equation}
We note that $dS=\xi R^2\cos\theta d\Omega$ with $\xi=+$ and $-$ for projection from the upper and lower halves of the Bloch sphere. By equaling Eqs.~(\ref{gm1}) and (\ref{gm2}) we obtain $B^\xi$ in Eq.~(\ref{magnetic}), which can also be calculated independently through the strain induced motion of the Dirac points. These two geometric phases were the two examples given in the original paper of Berry \cite{berry1984}. Here we show that in our specific model the two geometric phases can be mapped into each other. In contrast to a pseudo-spin-$1/2$ model that is usually used for the two valleys \cite{xiao2007}, here the $K$ and $K^\prime$ points correspond to the two poles of the Bloch sphere of spin-$N/2$.

To investigate the valley Hall effect, we introduce the potential due to a weak electric field  
$H_V=-eE_x x-eE_y y$. Here the field is weak, i.e., $euE_x, euE_y\ll g$ such that the transition between different Landau levels (i.e., the Landau-Zener tunneling) is negligible. We rewrite $x$ and $y$ in the $d$ modes and drop the terms containing $d_0$ and $d_0^\dagger$, which would induce interband transitions. Then we obtain
$X= x-P_y/eB^\xi=uJ_x$ and $Y= y+P_x/eB^\xi=uJ_y$, which, containing no $d_0$ terms, are the ${x}$- and ${y}$-coordinates of the guiding center, describing the motion within a Landau level, with $P_x$ and $P_y$, which contain $d_0$ terms, denoting the mechanical momenta involving transition between Landau levels \cite{Goerbig2011}  (see Supplementary material). Consequently, the valley Hall effect can be calculated by the following Hamiltonian,
\begin{equation}
H_V\rightarrow-eu\mathbf{E}\cdot\mathbf{J}.
\end{equation}
The valley Hall response of the electronic wavefunction is represented by the precession of a pseudo-spin around a pseudo-magnetic field pointing in the $x$-$y$ plane. The static electric field in the Hall effect plays the role of a pseudo-magnetic field.~Here the pseudo-magnetic field acting on the pseudo-spin (as mapped from the electric field in the lattice) should not be confused with the effective magnetic field due to strain of the lattice.

In the valley Hall effect, the strain-induced effective magnetic fields and thus the Hall responses at the two valleys have opposite signs \cite{guinea2010}. We assume that the electric field is applied along the $-{x}$ direction, and the Hamiltonian is $euE_x J_x$. The Heisenberg equation of $J_y$ is
\begin{equation}
\dot{J}_y=\frac{1}{i\hbar}[J_y, euE_x  J_x]=-\frac{euE_x J_z}{\hbar}.
\label{jy}
\end{equation}
At the $K$ and $K^\prime$ point, $J_z=\pm N/2$, we obtain 
$\dot{Y}=u\dot{J}_y=\mp {E_x}/{B_0}$,
which is evidence that the Hall responses at the two valleys have opposite signs, as shown in Fig.~\ref{fig3} (a) at $t=0$ and $T/2$. The electronic wavefunction oscillates in the incircle and gets reflected at the Lifshitz transition edge due to band-gap opening.

We introduce Haldane NNN hopping terms $H_H$ \cite{haldane1988,Ho2017}, e.g., $i\kappa [\sqrt{p(q+1)}a^\dagger_{(p-1)(q+1)r}a_{pqr}-\sqrt{pq}b^\dagger_{(p-1)qr}b_{p(q-1)r}]+{\rm h.c.}$ with complex hopping coefficients in the Fock-state lattice,
\begin{equation}
H_H\rightarrow -i\kappa \sigma_z \sum_jc^\dagger_{j+1} c_j+h.c.=-2\sqrt{3}\kappa\sigma_z J_z,
\end{equation}
where $\kappa$ is a coupling coefficient. Therefore, the Haldane terms correspond to a pseudo-magnetic field in ${z}$ direction and its signs are opposite for the two sublattices. In the zeroth Landau level, only the $a$-sublattice is occupied and $\sigma_z=-1$. The dynamics of the electronic wavefunction is mapped to the precession of a pseudo-spin in a pseudo-magnetic field pointing in the ${z}$ direction.
A wavepacket prepared on the incircle makes chiral rotation around the center of the lattice, which is the key feature of the Haldane model, as shown in Fig.~\ref{fig3} (b).

In conclusion, we propose an exactly solvable model of a strained lattice whose hopping rates are derived from the transitions between the Fock states of a three-mode JC model. The states in the zeroth Landau level of the strain-induced effective magnetic field can be represented by a pseudo-spin. We demonstrate that the valley Hall current can be represented by the spin precession around an in-plane pseudo-magnetic field, while the chiral edge current in the Haldane model can be represented by the spin precession around a pseudo-magnetic field perpendicular to the lattice plane. The method can be generalized to various strained lattices in arbitrary dimensions by introducing more atomic levels or cavity modes. This work reveals deep connections between the topological phases in condensed matter physics and cavity quantum electrodynamics.

This work was supported by the National Key Research and
Development Program of China (2019YFA0308100 and
2018YFA0307200), the National Natural Science Foundation of
China (11934011 and 11874322), the Strategic Priority Research
Program of Chinese Academy of Sciences (XDB28000000),
and the Fundamental Research Funds for the Central Universities.
H.C. was supported by the China Postdoctoral Science
Foundation (2019M650134). R.B.L. was supported by Hong Kong Research Grants Council-General Research Fund Project 14304117.

\end{document}